\documentclass[aps,prx,amsmath,amssymb,floatfix,twocolumn]{revtex4-1}
\usepackage{parskip} 
\usepackage{tikz} 
\usepackage{graphicx}
\usepackage{subfigure}
\usepackage{subcaption}
\usepackage{amsmath}
\usepackage{bbold}
\usepackage{slashed}
\usepackage{amssymb}
\usepackage{listings}
\usepackage{braket}
\usepackage{array}
\usepackage[colorlinks]{hyperref}
\usepackage{subcaption}
\usepackage{float}
\usepackage[margin=1in]{geometry}

\usepackage{natbib}

\begin{document}
\preprint{APS/123-QED}

\title{BerryEasy: A GPU enabled python package for diagnosis of $n$th-order and spin-resolved topology in the presence of fields and effects}
\author{Alexander C. Tyner$^{1,2}$}
\email{alexander.tyner@su.se}

\affiliation{$^{1}$ Nordita, KTH Royal Institute of Technology and Stockholm University 106 91 Stockholm, Sweden}
\affiliation{$^2$ Department of Physics, University of Connecticut, Storrs, Connecticut 06269, USA}

\date{\today}

\begin{abstract} 
Multiple software packages currently exist for the computation of bulk topological invariants in both idealized tight-binding models and realistic Wannier tight-binding models derived from density functional theory. Currently, only one package is capable of computing nested Wilson loops and spin-resolved Wilson loops. These state-of-the-art techniques are vital for accurate analysis of band topology. In this paper we introduce BerryEasy, a python package harnessing the speed of graphical processing units to allow for efficient topological analysis of supercells in the presence of disorder and impurities. Moreover, the BerryEasy package has built-in functionality to accommodate use of realistic many-band tight-binding models derived from first-principles. 
\end{abstract}

\maketitle
\par 
\section{Introduction} 
Immense progress has been made towards the understanding and cataloguing of non-trivial band topology in real systems\cite{tang2019efficient,zhang2019catalogue,vergniory2019complete,tang2019comprehensive,xu2020high}. Crucial to this progress has been the development of powerful community codes for the construction and analysis of tight-binding models. These programs include Z2Pack\cite{Z2pack}, WannierTools\cite{WU2017}, Wannier90\cite{Pizzi2020}, PythTB\cite{PythTB}, Kwant\cite{groth2014kwant}, WannierBerri\cite{tsirkin2021high}, and Pybinding\cite{moldovan2020pybinding} among others. While sharing many features, each package has individual strengths. Until recently a common issue was that none of these packages supported built-in functionality for computing state-of-the-art topological diagnostics including the nested Wilson loop\cite{Benalcazar61,Schindlereaat0346} and the spin-resolved Wilson loop\cite{Prodan2009,Lin2022Spin}. Both computations are critical for a comprehensive analysis of band topology. Recently, an auxiliary code was constructed to work in conjunction with the PythTB package for the computation of these quantities\cite{Lin2022Spin}. However, as mentioned, each code has separate strengths. PythTB is a terrific option for investigating band topology, however attempting to account for the presence of disorder/impurities in supercells can be cumbersome. Furthermore, it is not an ideal program for investigating transport or edge spectral density of many-band models. These are situations in which the WannierTools package has excelled due to its speed advantage through use of Fortran. However, as a Fortran based program it is challenging to manipulate, and does not offer access to the spin-resolved Wilson loop or nested Wilson loop at the moment. 
\par 
This situation has motivated the development of a new auxiliary package, BerryEasy, which works in tandem with the PyBinding software for computation of topological properties. PyBinding is designed with a python front end and C++ back end, offering a balance between speed and ease of use. It is thus ideal for investigating the effects of external fields, disorder and impurities on transport and density of states, but to this point it has lacked functionality for diagnosing band topology. In this paper, we detail how BerryEasy offers a simple interface for the computation of the Wilson loop\cite{NM1997,NM2007,Coh2009,yu2011equivalent,Soluyanov2011,WannierRepSol,alexandradinata2014wilson,Taherinejad2014,Z2pack,bouhon2019wilson,bradlyn2019disconnected}, nested Wilson loop\cite{Benalcazar61,Schindlereaat0346}, and spin-resolved Wilson loop\cite{Prodan2009,Lin2022Spin} for models defined in PyBinding. {\color{black}Given its popularity, we note that BerryEasy can additionally be interfaced with Kwant, although not all functionalities available via PyBinding are available with Kwant at this point. For clarity we focus on interfacing of BerryEasy with PyBinding in this work. For documentation and a tutorial on how to interface BerryEasy with Kwant please see the accompanying webpage\cite{github}.}

\par
Importantly, BerryEasy can be run using a CPU or GPU. Operating the program on a GPU is made possible by CuPy\cite{cupy_learningsys2017} and decreases the computation time dramatically. This is significant as computation of topological invariants is generally expensive, particularly in supercells, due to the need for exact diagonalization of a discretized Hamiltonian at many points in reciprocal space. In the BerryEasy workflow, the Hamiltonian is rapidly built by PyBinding's C++ back-end and the subsequent analysis by BerryEasy on GPUs enjoys significant speed advantages. As a result it is possible to efficiently investigate the fate of band-topology upon introduction of disorder, impurities, external fields and other physical situations for which finite-size effects and disorder averages must be considered. Finally, we incorporate functionality for defining a PyBinding lattice using a Wannier tight-binding model created using the Wannier90 software package. This extends all functionalities in idealized models to realistic systems. 
\par 
{\color{black}
The remainder of this work is organized as follows. In Sec. \eqref{sec:WCCs}, Wannier center charges and their use to determine topological properties is defined. The numerical approach used by BerryEasy for their determination is also presented. In Sec. \eqref{sec:Nested}, nested Wannier center charges are defined along with their computation in BerryEasy. In Sec. \eqref{sec:SpinResolved}, spin-resolved Wannier center charges are defined along with details of their computation in BerryEasy. Finally, before concluding in Sec. \eqref{sec:Summary}, details of how to analyze a realistic tight-binding model produced by Wannier90 are provided in Sec. \eqref{sec:Wan90}. In each section multiple examples are provided and in select cases these examples are accompanied by code snippets. These code snippets are offered to illustrate the ease with which BerryEasy can be implemented and do not represent the full code required to produce the results shown in the main body. Tutorials containing all necessary code to reproduce the data and figures seen in this work are available on the accompanying webpage\cite{github}. }

\section{Wannier center charges}\label{sec:WCCs}
Analysis of Wannier center charges (WCCs) as calculated via Wilson loops has proven to be a fundamental building block in diagnosis of band topology\cite{NM1997,NM2007,Coh2009,yu2011equivalent,Soluyanov2011,WannierRepSol,alexandradinata2014wilson,Taherinejad2014,Z2pack,bouhon2019wilson,bradlyn2019disconnected}. For clarity, we provide a brief overview of the formalism for computation of WCCs in this section and provide examples for implementation of the computation in the BerryEasy package. Mathematically, the Wannier center charge for band $n$ {\color{black}of a one-dimensional system} is defined as\cite{Z2pack}, 
\begin{equation}\label{eq:WCCs}
\bar{x}_{n}=\frac{ia}{2\pi}\int_{-\pi/a}^{\pi/a}dk \bra{u_{nk}}\partial_{k}\ket{u_{nk}}=\frac{a}{2\pi}\int_{-\pi/a}^{\pi/a}dk \mathcal{A}_{n}(k)
\end{equation}
where $a$ is the lattice constant, $\ket{u_{n}(k)}$ is the Bloch wavefunction corresponding to band $n$ and $\mathcal{A}_{n}(k)$ is the Berry gauge connection. We note that this computation {\color{black}requires that periodic boundary conditions be enforced} and path-ordering must be explicitly enforced if the Berry gauge connection is non-Abelian. When the Berry gauge connection is Abelian, computation of the Wilson loop is equivalent to computation of the one-dimensional winding number which, in the Altland-Zirnbauer table\cite{RyuRMP,RevModPhys.83.1057,ryu2010topological,RyuLudwigPRB,PhysRevB.88.075142,PhysRevB.88.125129}, is $\mathbb{Z}$ classified for class AIII insulators, such as the famous Su-Schreiffer-Heeger model\cite{ssh1979}. If the Berry gauge connection is non-Abelian, as is the case for spinful insulators in class AII, the Wilson loop is instead $\mathbb{Z}_{2}$ valued.
{\color{black}
\subsection{Numerical Implementation:}

\par 
In the BerryEasy package Wannier center charges are computed by discretizing the integral in eq. \eqref{eq:WCCs} and explicitly imposing path-ordering. This is accomplished via the formula, 
\begin{equation}\label{eq:NumWCCs}
    e^{i2\pi \bar{x}_{n}}=F_{n,k+N\Delta k}...F_{n,k+\Delta k}F_{n,k},
\end{equation}
where $F_{n,k+N\Delta k}=\langle u_{nk+\Delta k}|u_{nk}\rangle$,  $\Delta k =2\pi/(aN)$ where $a$ is the lattice constant, and $\ket{u_{nk}}$ is the Bloch function of band $n$ at reciprocal space coordinate $k$. The accuracy of the integral is determined by the parameter, $N$. This is an integer value which determines the number of discretized segments used in computing the integral and is set by the user in BerryEasy. We emphasize that periodic boundary conditions \emph{must} be imposed along the path of integration, i.e. $\psi(k)=\psi(k+2\pi/a)$. 
\par 
When considering supercells, including those which incorporate disorder and impurities, the above implementation remains valid provided we work in the language of twisted boundary conditions (TBCs). For a one-dimensional supercell of size $L$ with a unit lattice constant, TBCs are implemented as $\psi(x)=e^{i\theta_{x}}\psi(x+L)$ where $0\leq \theta_{x} <2\pi$. Following Ref. \cite{zhang2013coupling}, the single particle wavefunctions can then be Fourier transformed such that they are a function of the discrete momenta, $k=2\pi n_{x}/L+\theta_{x}/L$, where $0 \leq n_{x} < L$. 
\par 
In BerryEasy, when interfacing with PyBinding we fix $\theta_{x}=0$ by setting perioidc boundary conditions with a period equivalent to the supercell size and compute eq. \eqref{eq:NumWCCs} as a function of $k=2\pi n_{x}/L$. As such, when working with a supercell of size $L$, the reciprocal lattice vectors must be properly defined as $2\pi \rightarrow 2\pi/L$. Upon making this definition, the same numerical algorithms are implemented to rapidly compute the Wannier charge centers for the supercell. 
}

\begin{figure}
    \centering
    \includegraphics[width=7cm]{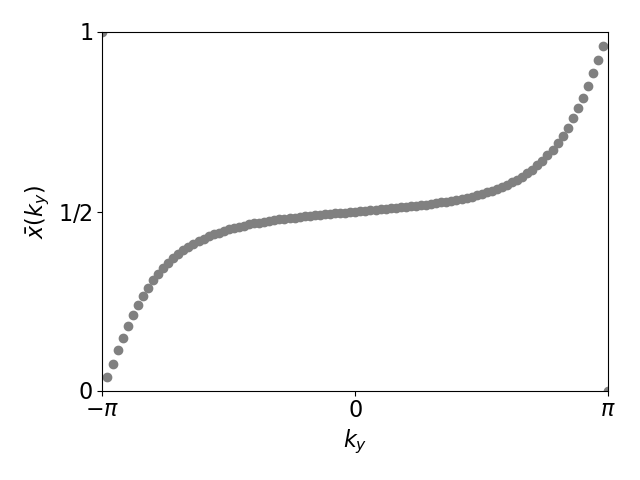}
    \caption{Wannier center charge (WCC) spectra for occupied valence band of eq. \eqref{eq:Chern}. Spectral flow of WCCs indicates a unit Chern number.}
    \label{fig:ChernWCC}
\end{figure}
\subsection{Hybrid Wannier center charges}\label{sec:HWCCs}
\par 
 Importantly, the WCC formalism has been extended to allow for the determination of bulk topological invarants in higher physical dimensions through computation of hyrbid WCCs. This is simply the computation of WCCs as a function of a transverse momenta. {\color{black} For example, in a two-dimensional system if we wish to compute the WCCs along $k_{x}$ as a function of transverse momenta $k_{y}$, we modify eq. \eqref{eq:WCCs} to the form, 
\begin{multline}\label{eq:HWCCs}
\bar{x}_{n}(k_{y})=\frac{ia}{2\pi}\int_{-\pi/a}^{\pi/a}dk_{x} \bra{u_{n\mathbf{k}}}\partial_{k_{x}}\ket{u_{n\mathbf{k}}}=\\ \frac{a}{2\pi}\int_{-\pi/a}^{\pi/a}dk_{x} \mathcal{A}_{n}(\mathbf{k}).
\end{multline}
We emphasize that in the above formula, both directions must support periodic boundary conditions. Indeed, to utilize WCCs for topological classification periodic boundary conditions must be imposed along all directions in the $d$-dimensional system we wish to classify. This is distinct from stating that periodic boundary conditions must be imposed along all directions. For example, if we wish to analyze a two-dimensional $x-y$ plane embedded in a three-dimensional system, the third direction ($z$) boundary conditions need not be periodic. 
\par
By analyzing spectral flow of the WCCs as a function of the transverse momenta, the presence or absence of topological invariants can be determined\cite{Z2pack}.} As an illustrative example, we consider a model of a Chern insulator on a square lattice\cite{Laughlin1981,TKNN1982,Niu1985,Haldane}. The Bloch Hamiltonian takes the form, 
\begin{equation}\label{eq:Chern}
H(\mathbf{k})/t= \sin k_{x}\sigma_{1}+\sin k_{y}\sigma_{2} + (\cos k_{x}+\cos k_{y}+ 1.5)\sigma_{3}
\end{equation}
where $t$ has units of energy, $\sigma_{j=1,2,3}$ correspond to the three Pauli matrices and the lattice constant has been set to unity. This model supports a non-trivial Chern number, $\mathcal{C}=1$ which can be computed as, 
\begin{equation}
    \mathcal{C}=\frac{1}{2\pi}\int_{BZ}d^{2}kF_{xy}(\mathbf{k}),
\end{equation}
where $F_{xy}(\mathbf{k})=\partial_{k_{x}}A_{y}-\partial_{k_{y}}A_{x}$ is the Abelian Berry curvature. Alternatively, the first Chern number can be computed using hybrid WCCs as, 
\begin{equation}\label{eq:WCCChern}
\mathcal{C}=\frac{1}{a}\left(\sum_{n}\bar{x}_{n}(k_{y}=2\pi)-\sum_{n}\bar{x}_{n}(k_{y}=0)\right),
\end{equation}
where the sum is over the occupied bands. Upon defining the Bloch Hamiltonian in the PyBinding package, we execute the calculation in the formalism of BerryEasy as,
\begin{widetext}
\begin{lstlisting}
import BerryEasy as be

vec=lambda t1, t2: [t1, t2, 0] #integation along kx as funct. of ky
ds=100 #discretizing integration into 100 steps
ds2=100 #discretize ky direction, 100 WCC computations will be performed
bands=[0] #Considering the occupied bands (can also use bands=range(1))
syst=chern_model #System we are considering, eq. 4, as PyBinding instance
rvec=2*np.pi*np.diag(np.ones(3)) #Reciprocal lattice vectors
WCCx=be.WSurf(vec,syst,bands,ds,ds2,rvec)
\end{lstlisting}
\end{widetext}
\par 
{\color{black}
We clarify that in the above code $vec=lambda\; t1,\; t2:\; [t1, t2, 0]$ is a Python lambda function defined by the user for determining the two-dimensional plane in which to perform computation of WCCs. Namely, $t1$ specifies the direction of integration in units of the reciprocal lattice vectors and $t2$ specifies the direction of the transverse momenta which will be varied to search for spectral flow of the WCCs. Both directions must be fixed to support periodic boundary conditions within PyBinding. In addition, the $rvec$ term specifies the reciprocal lattice vectors and must be a $3 \times 3$ array. This is the case even if the system is one or two-dimensional. In one dimensional systems the second and third row are discarded by BerryEasy and in two-dimensional systems the third row is discarded. 
\par
For clarity the above sample code shows only the final step in computing the WCCs via BerryEasy. In addition, the tight-binding model, eq. \eqref{eq:Chern}, must be defined in the formalism of PyBinding. In the above code this is represented by the $syst$ parameter. Further details of the code and data visualization for this example as well as all subsequent examples are available on the accompanying BerryEasy website\cite{github}. } The results of the computation, shown in Fig. \eqref{fig:ChernWCC}, detail that the WCCs smoothly interpolate between $0$ and $1$ as a function of $k_{y}$ indicating the Chern number $\mathcal{C}=1$.

\par
It is now important to consider a spinful system supporting time-reversal symmetry $\mathcal{T}$ with $\mathcal{T}^2=-1$. In this case the system belongs to class AII and supports a $\mathbb{Z}_{2}$ topological invariant under the ten-fold classification scheme\cite{RyuRMP}. Importantly, it was shown that a direct connection can be made between the $\mathbb{Z}_{2}$ index\cite{FuKane,FKPump} and the WCC spectra in Ref. \cite{yu2011equivalent}. We briefly summarize that in a time-reversal symmetric system each Kramers pair is composed of two eigenstates which admit equal and opposite Chern numbers, $\mathcal{C}_{1}=-\mathcal{C}_{2}$. The $\mathbb{Z}_{2}$ index is then determined as, $\nu =(\mathcal{C}_{1}-\mathcal{C}_{2})/2 \;\text{Mod} \; 2$. As a result, a non-trivial $\mathbb{Z}_{2}$ index is indicated by the presence of a WCC spectra simultaneously detailing a smooth interpolation from $\bar{x}(0)=0(a)$ to $\bar{x}=a(0)$. 
\begin{figure}
    \centering
    \includegraphics[width=7cm]{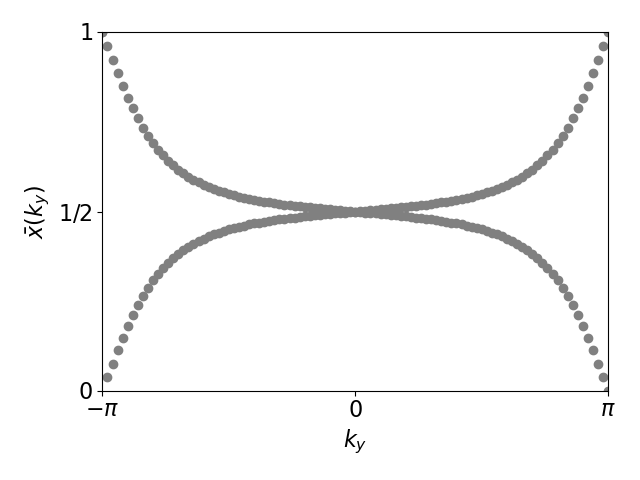}
    \caption{Wannier center charge (WCC) spectra for occupied valence band of eq. \eqref{eq:Model}. Gapless nature of WCC spectra indicates a non-trivial $\mathbb{Z}_{2}$ index.}
    \label{fig:BHZWCC}
\end{figure}
\par 
As an example of this behavior we will modify the Bloch Hamiltonian in eq. \eqref{eq:Chern} to the form, 
\begin{multline}\label{eq:Model}
    H(\mathbf{k})/t=\sin k_{x} \sigma_{1}\otimes \tau_{1} + \sin k_{y}\sigma_{1}\otimes \tau_{2} \\+ (\cos k_{x}+\cos k_{y}+ 1.5)\sigma_{3}\otimes \tau_{0} ,
\end{multline}
where $\sigma_{0,1,2,3}(\tau_{0,1,2,3})$ are the $2\times 2$ identity matrix and three Pauli matrices respectively, operating on the spin (orbital) indices. This is the celebrated Bernevig-Hughes-Zhang model\cite{bernevig2006quantum,ZhangFT} of the quantum spin-Hall insulator. Using the BerryEasy package, the WCC spectra is then computed as, 
\begin{widetext}
\begin{lstlisting}
import BerryEasy as be

vec=lambda t1, t2: [t1, t2, 0] #integation along kx as funct. of ky
ds=100 #discretizing integration into 100 steps
ds2=100 #discretize ky direction, 100 WCC computations will be performed
bands=[0,1] #Considering the occupied bands (can also use bands=range(2))
syst=BHZ_model #System we are considering, eq. 7, as PyBinding instance 
rvec=2*np.pi*np.diag(np.ones(3)) # Reciprocal lattice vectors
WCC=be.WSurf(vec,syst,bands,ds,ds2,rvec)
\end{lstlisting}
\end{widetext}

{\color{black}Again, details of the code for defining eq. \eqref{eq:Model} as a PyBinding instance are available in the tutorial on the accompanying webpage\cite{github}.} The results in Fig. \eqref{fig:BHZWCC} demonstrate the resulting fully connected WCC spectra indicating a non-trivial $\mathbb{Z}_{2}$ index. This formalism can be easily extended to three-dimensional systems as computation of the weak and strong $\mathbb{Z}_{2}$ indices is accomplished via computation of the $\mathbb{Z}_{2}$ index in each high-symmetry plane\cite{FuKane}.

\begin{figure*}[t]
\subfigure[]{
\includegraphics[scale=0.35]{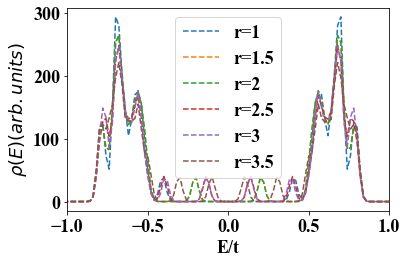}
\label{fig:dosvac}}
\subfigure[]{
\includegraphics[scale=0.35]{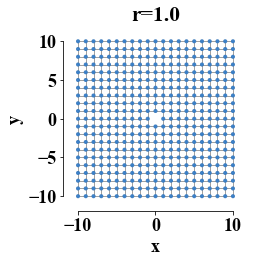}
\label{fig:r1}}
\subfigure[]{
\includegraphics[scale=0.35]{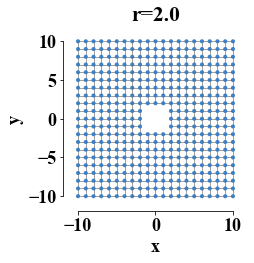}
\label{fig:r2}}
\subfigure[]{
\includegraphics[scale=0.35]{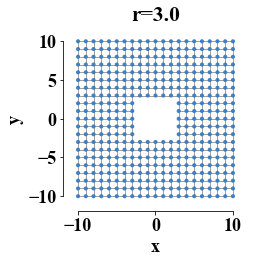}
\label{fig:r3}}
\subfigure[]{
\includegraphics[scale=0.33]{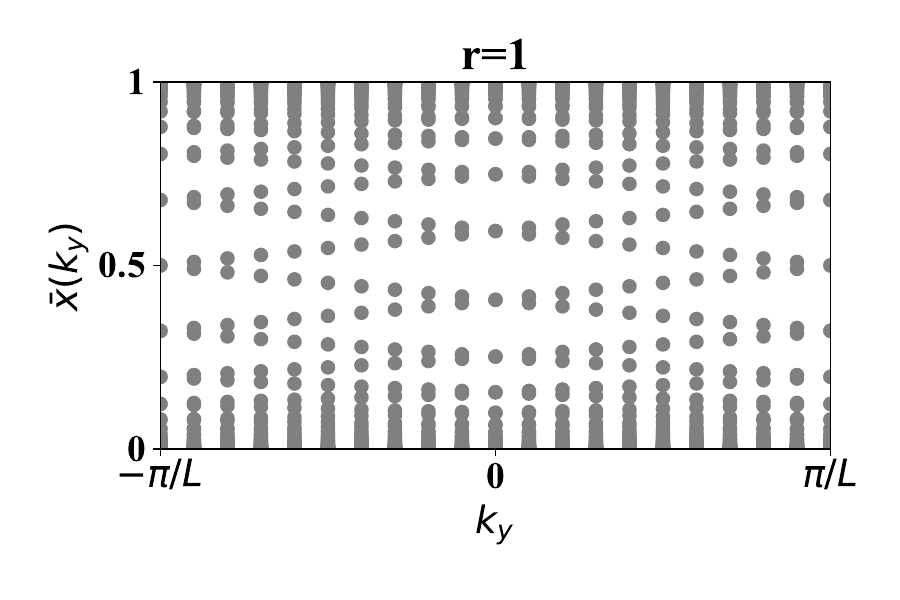}
\label{fig:WCCr1}}
\subfigure[]{
\includegraphics[scale=0.33]{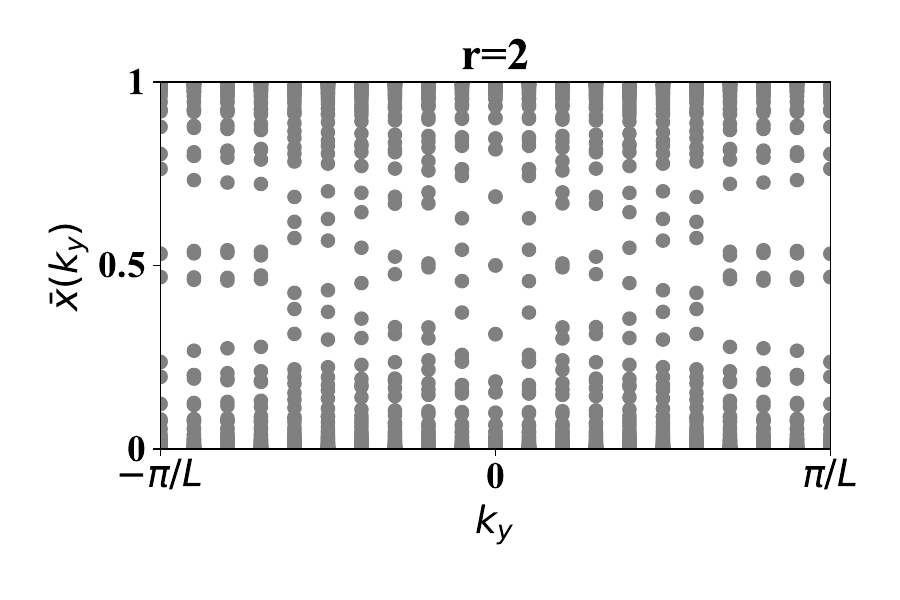}
\label{fig:WCCr2}}
\subfigure[]{
\includegraphics[scale=0.33]{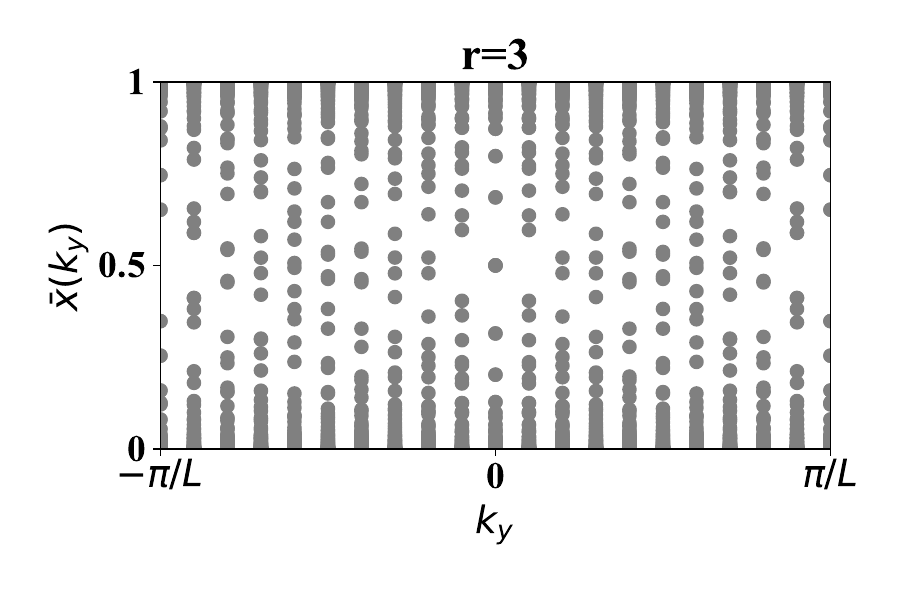}
\label{fig:WCCr3}}
\caption{(a) Density of states for $21\times 21$ supercell as a function of vacancy size under periodic boundary conditions. (b)-(d) Supercell with vacancies at origin within a circle of radius $r$. (e)-(g) Wannier center charge spectra for supercells shown in (b)-(d) respectively.}
\label{fig:Defect}
\end{figure*}

\subsection{Example: Topological Insulator with vacancies}\label{sec:Vacancy}
\par
As discussed in the introduction, a benefit of the PyBinding package is the capability to include fields and defects with ease. Accounting for defects is critical to the investigation of any realistic experimental setup but computing topological invariants in the presence of defects such as vacancies can be exceedingly challenging. In this example, we consider the BHZ model of a two-dimensional quantum spin-Hall insulator with a non-trivial $\mathbb{Z}_{2}$ index\cite{bernevig2006quantum,ZhangFT}. We will create a supercell of $21 \times 21$ unit cells. We then remove lattice sites contained within a radius $r$ centered at the origin. As $r$ is increased the $\mathbb{Z}_{2}$ index will be computed. For a Jupyter notebook containing full details of this example please consult Ref. \cite{github}. 
\par
Vacancies can be included in lattice tight-binding models simply in the PyBinding package through the use of the site-state modifier function, {\color{black}the complete code to construct the model is given on the webpage\cite{github}}. As stated we will consider a $21 \times 21$ supercell of the model given in eq. \eqref{eq:Model}. A plot of the supercell and bulk density of states for varying values of $r$ is shown in Fig. \eqref{fig:dosvac} as well as images of the supercell with vacancies in Figs. \eqref{fig:r1}-\eqref{fig:r3}. We note as $r$ is increased the density of states within the bulk gap begins to become populated. This can be understood in a straightforward manner: increasing $r$ causes the vacancies to appear as a boundary with edge states. The limited size of the boundary means the hybridization of these edge states maintains an energetic gap, protecting the bulk topology. 
\par 
To test this, the $\mathbb{Z}_{2}$ index can be computed via the Wannier center charge spectra of the supercell in the presence of the vacancies. The results of this computation are available in Figs. \eqref{fig:WCCr1}-\eqref{fig:WCCr3}, indicating that indeed the spectra remains gapless and the $\mathbb{Z}_{2}$ index is intact.

\section{Nested WCC spectra}\label{sec:Nested}
There are currently limited community codes which accommodate computation of the nested Wilson loop. The nested Wilson loop or nested WCC spectra was introduced in Refs. \cite{Benalcazar61,BenalacazarCn,Schindlereaat0346} as a method for computing the WCC spectra of the Wannier Hamiltonian. As an example, the Wannier Hamiltonian for a two-dimensional system takes the form, $H_{W1}(k_{2})$, computed as, 
\begin{equation}\label{eq:WannierHam}
    H_{W1}(k_{2})=\text{Im}\left(\text{Ln}\left(\mathcal{P}\text{exp}\left( i \int_{-\pi/a}^{\pi/a}dk A_{1}(k_{2})\right) \right)\right),
\end{equation}
where $\mathcal{P}$ indicates path-ordering and $A_{1}(k_{2})$ is the Berry gauge connection for occupied bands. {\color{black} We therefore note that the Wannier Hamiltonian can be constructed numerically in BerryEasy by replacing the right hand side of eq. \eqref{eq:WannierHam} with the dicretized formulation of the integral in eq. \eqref{eq:NumWCCs}. It is also important to  emphasize that this computation requires periodic boundary conditions are imposed along $k_{1}$ and $k_{2}$.} 
\par
\begin{figure*}
    \subfigure[]{
    \includegraphics[width=5cm]{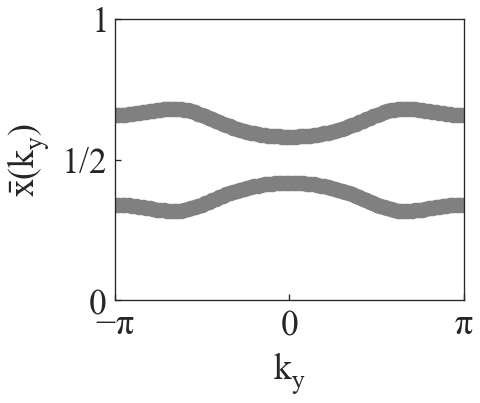}
    \label{fig:CHWCC}}
    \subfigure[]{
    \includegraphics[width=5cm]{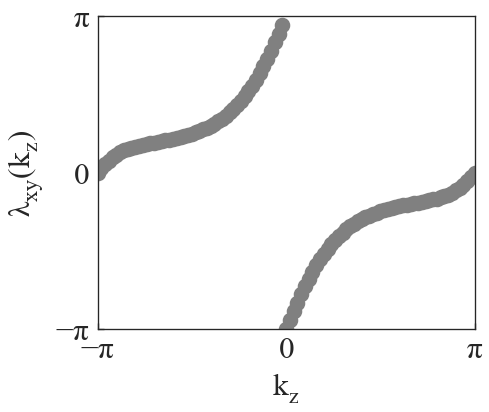}
    \label{fig:NWL}}
    \subfigure[]{
    \includegraphics[width=5cm]{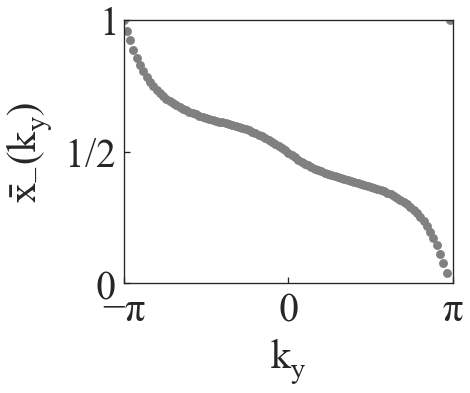}
    \label{fig:CHSpinWCC}}
    \caption{(a) Gapped Wannier center charge spectra of eq. \eqref{eq:CHModel} at $k_{z}=0$ plane. (b) Spectra as a function of $k_{z}$ of Wannier Hamiltonian constructed via Wilson loop along the $k_{x}$ direction as a function of $k_{y}$, also known as nested Wilson loop. (c) Spin-resolved Wannier center spectra detailing presence of non-trivial spin-Chern number in the $k_{z}=0$ plane of eq. \eqref{eq:CHModel}.}
    \label{fig:NL}
\end{figure*}

 
While claims that the Wannier Hamiltonian is equivalent to the surface Hamiltonian have been made to justify the use of the nested Wilson loop in diagnosis of higher-order topological insulators where the surfaces can take the form of lower dimensional topological insulators, this remains unresolved with counter examples appearing in the literature\cite{PhysRevB.107.075126}. Nevertheless, the nested Wilson loop has proven incredibly useful and powerful in identifying higher-order topological insulators (HOTIs). While the nested Wilson loop formalism is provided in detail in Ref. \cite{Benalcazar61}, here we provide an example for its implementation using the BerryEasy package. For clarity we will utilize the well studied chiral HOTI Bloch Hamiltonian\cite{Schindlereaat0346} which takes the form, 

\begin{multline}\label{eq:CHModel}
    H(\mathbf{k})/t=\sin k_{x} \sigma_{1}\otimes \tau_{1} + \sin k_{y}\sigma_{1}\otimes \tau_{2}+\sin k_{z}\sigma_{1}\otimes \tau_{3} \\+ (\cos k_{x}+\cos k_{y}+ \cos k_{z} -1.5)\sigma_{3}\otimes \tau_{0}\\ + \Delta_{0}(\cos k_{x}-\cos k_{y})\sigma_{2}\otimes \tau_{0}.
\end{multline}
If $\Delta_{0}$ is set to zero the model reduces to that of a strong topological insulator. Furthermore, setting $\Delta_{0}=0$ and $k_{z}=0$ this model reduces to the BHZ two-dimensional QSH insulator investigated previously. However, by setting $\Delta_{0}=1$, the surface states are gapped on the $(100)$ and $(010)$ surfaces. As a result, the WCC spectra is gapped, as seen in Fig. \eqref{fig:CHWCC} for the $k_{z}=0$ plane, disallowing topological classification as a strong topological insulator. At this point we compute the nested WCC spectra as a function of $k_{z}$. In the BerryEasy package this performed as,

\begin{widetext}
\begin{lstlisting}
import BerryEasy as be
NL=[]
for kz in np.linspace(0,1,21):
    vec=lambda t1, t2: [t1, t2, kz] #integation along kx as funct. of ky at kz=0
    ds=100 #discretizing integration along kx into 100 steps
    ds2=100 #discretize integration along ky into 100 steps
    bands=[0,1] #Considering the occupied bands (can also use bands=range(2))
    syst=CHOTI_model #System we are considering, eq. 9, as PyBinding instance 
    rvec=2*np.pi*np.diag(np.ones(3)) # Reciprocal lattice vectors
    NL.append(be.WNWL(vec,syst,bands,ds,ds2,rvec))
\end{lstlisting}
\end{widetext}

{\color{black}
The above code example represents the final step in computing the nested Wilson loop, the code to define eq. \eqref{eq:CHModel} in the format of PyBinding, i.e. the $syst$ parameter, is available on the corresponding webpage\cite{github}.}

\par 
The results seen in Fig. \eqref{fig:NWL}, demonstrate that the nested Wilson loop shows a gapless spectra indicating that the surface resembles a non-trivial Chern insulator with gapless chiral edge states. These edge states are the famous chiral hinge modes.

\section{Spin-resolved WCC spectra}\label{sec:SpinResolved}
While first introduced by Prodan\cite{Prodan2009} over a decade previously, the spin-resolved Chern number ($\mathcal{C}_{s}$) has found renewed importance in the diagnosis of band topology for spinful higher-order and fragile topological insulators as the crystalline symmetry preserving perturbations which serve to gap the edge states and bring about higher-order topology often violate the spin-rotation symmetry, forcing the use of this method to diagnosis the bulk invariant. The spin-resolved WCC spectra is detailed at length in Ref. \cite{Prodan2009,Lin2022Spin}, where a python package for its computation within the PythTB package is provided. This is a terrific community code. Nevertheless, we have found a need for implementation of the spin-resolved Wilson loop in PyBinding given the ease with which realistic fields and perturbations can be included. As an example, we again utilize the Bloch Hamiltonian given in eq. \eqref{eq:CHModel}. As seen in Fig. \eqref{fig:CHWCC}, the WCC spectra is gapped. Computation of the spin-resolved WCC spectra requires defining the projected spin operator (PSO), $P(\mathbf{k})\hat{s}P(\mathbf{k})$, where $P(\mathbf{k})$ is the projector onto occupied bands and $\hat{s}$ is a chosen spin-quantization axis. In the absence of spin-orbit coupling the eigenvalues of the PSO are fixed as $+ 1$ and $-1$, corresponding to the spin-up and spin-down eigenstates respectively. Since we have introduced spin-orbit coupling in eq. \eqref{eq:CHModel}, the eigenvalues can adiabatically deviate from $\pm 1$. Nevertheless, a gap in the eigenvalue spectra of the PSO remains when selecting $\hat{s}=s_{z}=\sigma_{3}\otimes \tau_{3}$, allowing for calculation of the WCC spectra for the negative eigenstates of the PSO, $\bar{x}_{-}(k_{y})$, as detailed in Lin et. al\cite{Lin2022Spin}. Such a calculation directly reveals $\mathcal{C}_{\downarrow}$. The results of this calculation fixing $k_{z}=0$ are seen in Fig. \eqref{fig:CHSpinWCC}, detailing the presence of a non-trivial spin-Chern number, $|\mathcal{C}_{s}|=1$.

This computation is performed in the BerryEasy package via the following lines of code:
\begin{widetext}
\begin{lstlisting}
import BerryEasy as be

vec=lambda t1, t2: [t1, t2, 0] #integation along kx as funct. of ky at kz=0
ds=100 #discretizing integration into 100 steps
ds2=100 #discretize ky direction, 100 WCC computations will be performed
bands=[0,1] #Considering the occupied bands (can also use bands=range(2))
syst=CHOTI_model #System we are considering, eq. 9, as PyBinding instance
rvec=2*np.pi*np.diag(np.ones(3)) # Reciprocal lattice vectors
op=np.diag([1,-1,-1,1]) #Array defining the prefered spin axis operator
WS=be.WSpinSurf(vec,model,bnds,ds,ds2,op,rvec)

\end{lstlisting}
\end{widetext}

\begin{figure*}
    \centering
    \includegraphics[width=16cm]{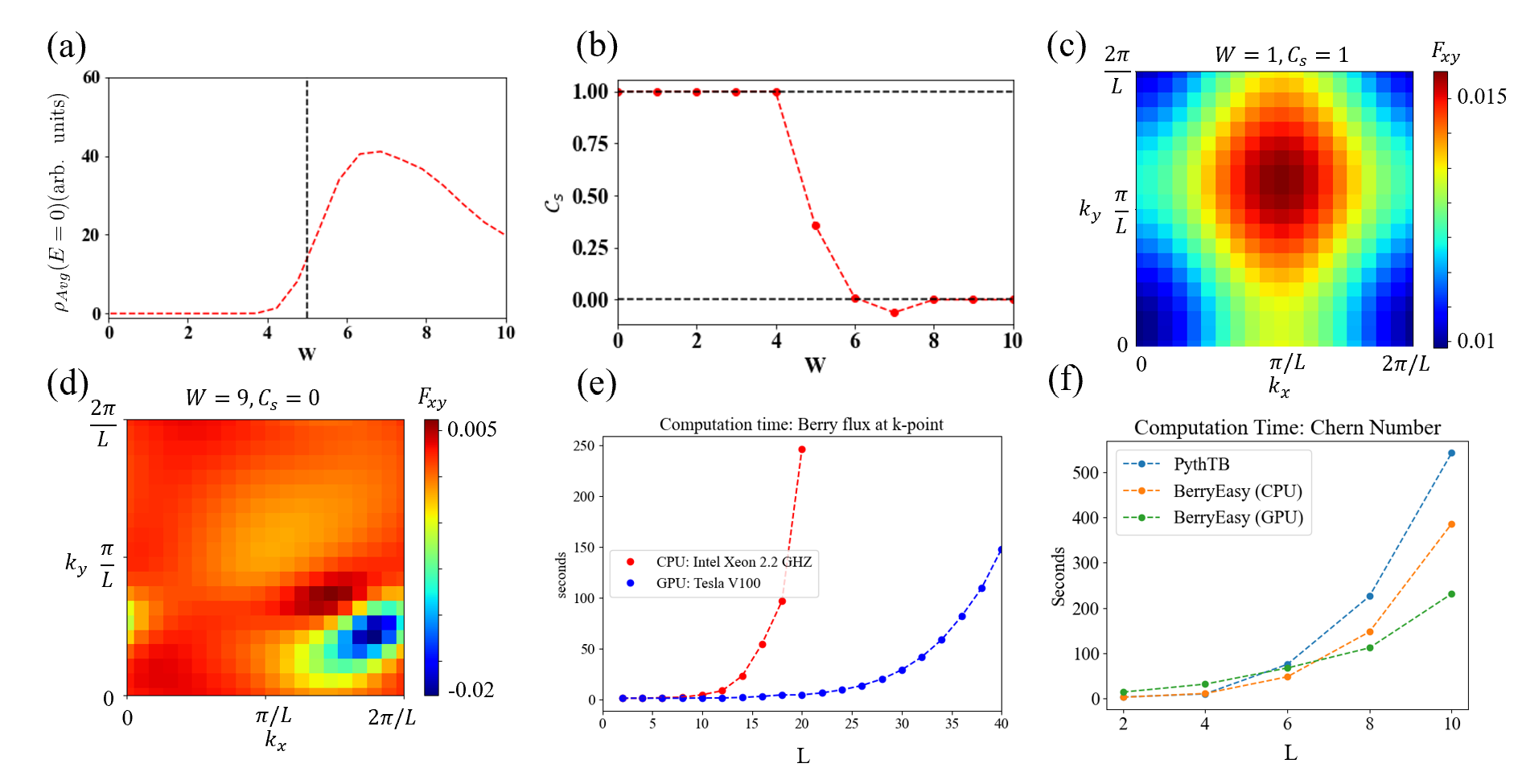}
    \caption{(a) Average density of states at zero energy for eq. \eqref{eq:quadrupolar} upon introduction of chiral symmetry preserving disorder in a $100 \times 100$ system with periodic boundary conditions. (b) Spin-chern number of occupied ground-state as a function of disorder strength, averaging over 20 disorder configurations for a $21\times21$ supercell. (c)-(d) Momentum resolved Berry flux for (c) $W=1$ and (d) $W=9$ yielding $\mathcal{C}_{s}=1$ and $\mathcal{C}_{s}=0$ respectively. (e) Comparison of time for computation of Berry flux at a single $k$-point in momentum space utilizing a CPU or GPU as a function of the supercell dimensions. (f) Comparison of computation time in seconds to compute the Chern number of eq. \eqref{eq:Chern} to identical accuracy in PythTB and BerryEasy in a supercell of linear system size $L$.}
\label{fig:Disorder}
\end{figure*}

\subsection{Example: Disordered quadrupolar insulator} \label{sec:QuadrupoleDisorder}
The study of disorder in topological quantum matter is of supreme importance yet it presents a computational challenge\cite{Haldane,Sheng2006,Onoda2007,Obuse2008,Prodan2010}. 
In order to demonstrate the benefits of the BerryEasy program and its integration with the PyBinding package, we consider the case of a disordered quadrupolar insulator\cite{Benalcazar61}. Disorder is generally difficult to implement in alternative packages, however, PyBinding's built in on-site modifiers combined with the BerryEasy package allow for an efficient route to establishing bulk topological invariants in disordered systems. We will consider a spinful version of the celebrated Benalcazar-Bernevig-Hughes model\cite{Benalcazar61}, the Bloch Hamiltonian is given as, 
\begin{multline}\label{eq:quadrupolar}
    H(\mathbf{k})/t=\sin k_{x} \sigma_{1}\otimes \tau_{3} + \sin k_{y}\sigma_{2}\otimes \tau_{3} \\+ (\cos k_{x}+\cos k_{y}+ 1)\sigma_{3}\otimes \tau_{3} \\+ 0.5(\cos k_{x}-\cos k_{y})\sigma_{0}\otimes \tau_{0}.
\end{multline}
This Hamiltonian supports chiral symmetry, generated as $S^{-1}HS=-H$, where $S=\sigma_{0}\otimes \tau_{2}$. In Refs. \cite{Li2020,Hu2021,Yang2021}, the robustness of the bulk topology to chiral symmetry preserving disorder was studied in depth. It was shown that the bulk spin-Chern number, as defined using the method of Prodan\cite{Prodan2009} fixing $\hat{s}=\sigma_{0}\otimes \tau_{3}$ in the PSO, was robust prior to the closing of the average bulk energetic gap. Following Ref. \cite{Hu2021}, this disorder will be implemented as $H_{dis}=\{w_{1},w_{1}\}\otimes \tau_{3}$, where $w_{j}$ is selected randomly from a uniform distribution, $[-W/2,W/2]$. The bulk average density of states can be calculated rapidly for a $100 \times 100$ system using the built-in Kernel Polynomial Method (KPM) solvers in PyBinding, averaging over 100 disorder configurations (for a Jupyter Notebook detailing this example please consult Ref. \cite{github}). The results shown in Fig. \eqref{fig:Disorder}(a) demonstrate that the average bulk gap remains open until $W\approx 5$. In order to determine the spin-Chern number in the presence of disorder we employ the coupling matrix approach. This is implemented by first creating a supercell of size $21 \times 21$ lattice cells and applying the chiral-symmetry preserving disorder. We then impose periodic boundary conditions such that the $21 \times 21$ system represents the fundamental unit cell and the reciprocal lattice vectors are altered as $2\pi \rightarrow 2\pi/(21a)$, where $a$ is the lattice constant which we set to unity. As we are working with a supercell, the reciprocal lattice vectors and the number of valence bands must be modified to account for the increase in the dimensions of the Bloch Hamiltonian. We then perform the spin-Chern number computation. This is accomplished utilizing the $WSpinLine$ function in BerryEasy, which allows for explicit construction of closed lines in momentum space along which the Wilson loop is computed. These closed paths are constructed by discretizing the Brillouin zone into a grid of $20\times 20$ plaquettes for which the integrated Berry flux is computed. This represents a spin-resolved generalization of the coupling matrix method introduced in Ref. \cite{zhang2013coupling}. The spin-Chern number is then averaged over 20 disorder configurations. The results shown in Fig. \eqref{fig:Disorder}(b), demonstrate that the spin-Chern number remains intact prior to closing of the average bulk gap. A sample output of this computation in the region before and after closing of the bulk gap is shown in Figs. \eqref{fig:Disorder}(c) and \eqref{fig:Disorder}(d) respectively.   
\par 
\subsubsection{GPU utilization to expedite calculations:} \label{sec:GPU}
The effects of disorder in solid-state systems admitting non-trivial band topology continues to be an area of active research. In general, the bulk topological invariant is robust to the introduction of disorder which leaves the bulk mobility gap intact. However, measurement of the bulk topological invariant in disordered systems is known to be a computationally demanding endeavor due to the need to simultaneously avoid finite size effects and average over many disorder configurations. 
\par 
In this example, we have showcased the ability of BerryEasy package to compute the spin-resolved Chern number, a capability offered by only one other community code, and to perform the computation rapidly in a large supercell accounting for disorder utilizing the power of GPUs. In Fig. \eqref{fig:Disorder}(e), we plot the time for computing integrated Berry flux for the occupied states of eq. \eqref{eq:quadrupolar} through a single plaquette in an $L\times L$ supercell of the BerryEasy package when utilizing a Tesla V100 GPU vs Intel-Xeon CPU as offered in Google Colab.
\par
{\color{black}\emph{Comparison with PythTB:}  To further demonstrate the speed advantage of BerryEasy, we plot the total time required to compute the Chern number for eq. \eqref{eq:Chern} on a supercell of linear system size $L$ using BerryEasy and PythTB to an identical accuracy in Fig. \eqref{fig:Disorder}(f). The results demonstrate the enormous advantage of BerryEasy, particularly when run on a GPU. This speed advantage is of vital importance to achieve accurate results as finite size effects can be minimized and a greater number of disorder configurations can be considered without requiring extended computational time. The code utilized for the results in Fig. \eqref{fig:Disorder} is available online\cite{github}.  }
\par
{\color{black}
\emph{GPU environment requirements:} The GPU enabled version of BerryEasy relies on usage of the CuPy package\cite{cupy_learningsys2017}. As a result, requirements for GPU environment are the same as those for use of CuPy. Namely, CuPy requires a NVIDIA CUDA GPU with CUDA Toolkit V11.2 or higher.  }

\section{Wannier90 tight-binding models}\label{sec:Wan90}
Finally, progress in the diagnosis of band topology has been accelerated by the ability to directly utilize the tools listed above in realistic tight-binding models derived from density functional theory using the Wannier90 software package\cite{Pizzi2020}. A primary advantage of the PythTB, Z2Pack and WannierTools software packages is their ability to directly interface with the output of Wannier90 for construction of a tight-binding model from which topological quantities can be computed using the built in functionalities. The BerryEasy package provides a built-in version of wanPB\cite{wanPB} to create PyBinding instances from the output of Wannier90, thereby extending the functionality of the tools detailed above to realistic many-band models generated from density functional theory.

\begin{figure*}[t]
\centering
\subfigure[]{
\includegraphics[scale=0.6]{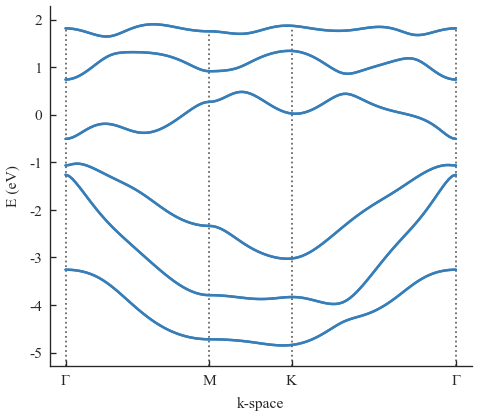}
\label{fig:BiBands}}
\subfigure[]{
\includegraphics[scale=0.5]{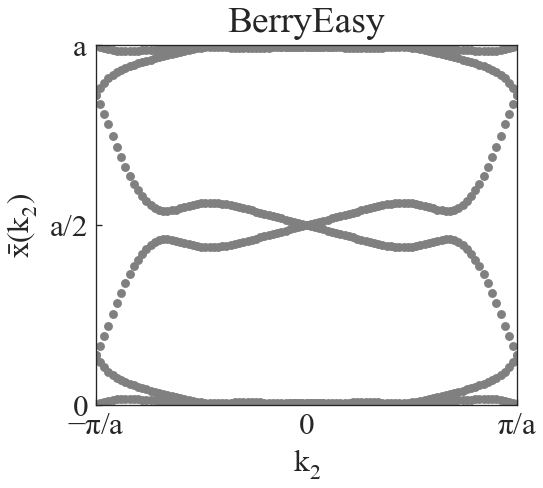}
\label{fig:BEWCCBi}}
\subfigure[]{
\includegraphics[scale=0.5]{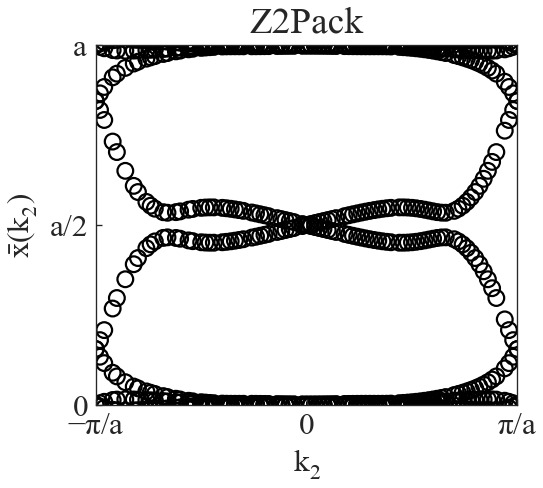}
\label{fig:Z2WCCBi}}
\caption{(a) Band structure of bilayer-bismuth along high-symmetry path in the Brillouin zone. The Wannier center charge spectra is computed for the three lowest lying bands using both the (a) BerryEasy and (c) Z2Pack software package detailing identical results which classify the system as $\mathbb{Z}_{2}$ non-trivial. }
\label{fig:Bi}
\end{figure*}

\par 
As an example, we examine a known $\mathbb{Z}_{2}$ topological insulator in two-dimensions, bilayer-bismuth\cite{WadaBi,MurakamiQSHBi,tynerbismuthene}, also known as $\beta$-bismuthene. {\color{black}An additional three-dimensional example, the topological Dirac semimetal PdTe$_{2}$\cite{pdte2}, is given in the tutorial on the webpage\cite{github}.} All first principles calculations based on density-functional theory (DFT) are carried out using the Quantum Espresso software package \cite{QE-2009,QE-2017,QE-2020}. Exchange-correlation potentials use the Perdew-Burke-Ernzerhof (PBE) parametrization of the generalized gradient approximation (GGA) \cite{Perdew1996}. The self-consistent and non-self consistent computations are performed using a $40 \times 40 \times 1$ Monkhorst-Pack grid and a cutoff of 100 Ry. We implement norm-conserving pseudo-potentials\cite{Hamann2013} as obtained on the Pseudo-Dojo site\cite{van2018pseudodojo}. Spin-orbit coupling is consider in all calculations. The Wannier tight-binding model was constructed using the Wannier90 software package. The necessary input and output files are available publicly\cite{github}. Utilizing the Bi\_tb.dat and Bi\_centres.xyz files, we create the PyBinding tight-binding model using the following function to create the lattice and save it,

\begin{lstlisting}
import BerryEasy as be

ECut=0.05 #cutoff for hopping strength 
HopCut=100 #hopping neighbors cutoff
seedname=str(Bi)
lat = be.wan90_lat(ECut,HopCut,seedname)
pb.save(lat,"BiWTB.pbz")
\end{lstlisting}

Upon plotting the band structure seen in Fig. \eqref{fig:BiBands}, we investigate the nature of the ground state topology using the built-in function for computing WCCs. Besides specifying the correct number of valence bands and the correct reciprocal lattice vectors, the necessary code is virtually unchanged from that displayed for eq. \eqref{eq:Model}. The results are then plotted showing that the hybrid WCC spectra is gapless in Fig. \eqref{fig:BEWCCBi}. As such the system can be classified as a strong topological insulator. While such a computation can be performed using any of the software packages given previously (this is demonstrated by comparing the results to those obtained by the Z2Pack software package to show that they are identical in Fig. \eqref{fig:Z2WCCBi}), we remark that the ability to implement perturbations such as disorder or impurities now remains possible even in the context of complicated Wannier tight-binding models through the interface with PyBinding.  
\par
\begin{figure}
\centering
\subfigure[]{
\includegraphics[width=3.5cm]{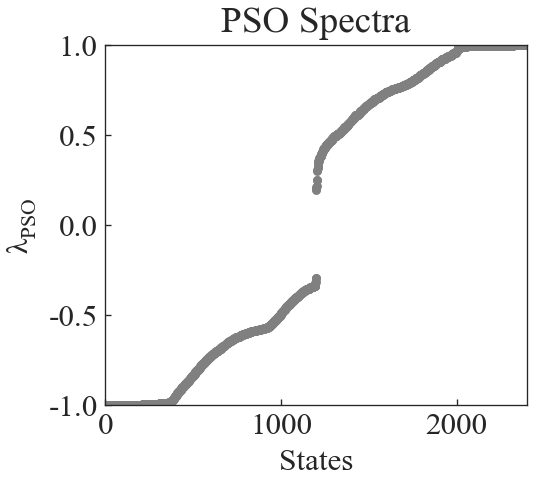}
\label{fig:PSOSpectra}}
\subfigure[]{
\includegraphics[width=3.7cm]{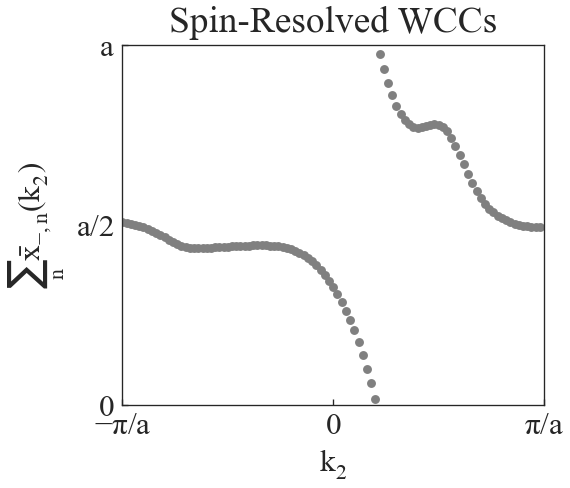}
\label{fig:PSOWCC}}

\caption{(a) Eigenvalues of the projected-spin operator when sampling the Brillouon zone along a $20 \times 20$ grid. The results detail the presence of a spectral gap allowing for topological classification of the negative(positive) eigenstates to determine $\mathcal{C}_{\downarrow}(\mathcal{C}_{\uparrow})$. (b) Sum of the Wannier center charges for the negative eigenstates of the projected spin-operator at each value of $k_{2}$. The spectral flow allows for determination of $\mathcal{C}_{\downarrow}=-1$ utilizing eq. \eqref{eq:WCCChern}.}
\label{fig:BiSpin}
\end{figure}

 
{\color{black}
Having determined that $\beta$-bismuthene supports a non-trivial $\mathbb{Z}_{2}$ index, it is important to determine the magnitude of the spin-Chern number as in the presence of a non-trivial $\mathbb{Z}_{2}$ index, it has been shown that $\nu_{0}=|\mathcal{C}_{s}|\text{Mod}2$. As a result, for $\beta$-bismuthene, previous works have predicted $\mathcal{C}_{s}=\pm 1$ or $\mathcal{C}_{s}=\pm 3$\cite{MurakamiQSHBi}, but been unable to distinguish between the two classifications. Here we show that the spin-Chern number can be directly computed for $\beta$-bismuthene using BerryEasy. To do so it is first necessary to determine the preferred spin-direction for which the PSO is fully gapped. This is accomplished within the BerryEasy package using the $spin\_ spectrum $ function. This function takes as parameters the $k$-point at which to compute the spectra of the PSO, the PyBinding instance, the bands from which to form the PSO, and the spin operator. Through a trial and error process we have identified the proper spin-operator which produces a gapped PSO for the ground-state of $\beta$-bismuthene. This is accomplished by discretizing the Brillouin zone into a grid of  400 $k$-points, collecting the eigenvalues of the PSO at each point on the grid, and plotting the eigenvalues to identify the presence of a spectral gap. The results in Fig. \eqref{fig:PSOSpectra} clearly demonstrate that the PSO is gapped for the chosen spin operator. 
\par 
Having identified the proper spin-operator, we compute the spin-resolved WCC spectra in a manner identical to that performed for eq. \eqref{eq:CHModel}. To determine the spin-Chern number we then invoke eq. \eqref{eq:WCCChern}, plotting $\sum_{n}\bar{x}_{-,n}(k_{2})$ in Fig. \eqref{fig:PSOWCC}. The results clearly demonstrate that the system supports $\mathcal{C}_{s}=+1$, resolving the ambiguity posed in earlier studies of the system. }

\section{Summary}\label{sec:Summary}
\par
In summary, multiple excellent community codes currently exist for the computation of topological quantities in both Wannier tight-binding models and idealized toy models. However, each package is ideally suited to different use cases. The BerryEasy package fills what we view as a current gap in this spectra of offerings by providing access to state-of-the-art diagnostics for toy models and realistic models while simultaneously offering the possibility to account for application of fields and effects by interfacing with the PyBinding package. As a result, it is our hope that this package will prove useful both for those researchers investigating new forms of topological order in clean systems as well as those interested in understanding the complex effects of disorder and application of external fields in realistic systems being actively studied in experimental settings. {\color{black}While BerryEasy is capable of interfacing with Kwant\cite{groth2014kwant}, future versions of the BerryEasy package are currently being developed to allow for all functions offered via PyBinding. Future versions are further being developed to accommodate tight-binding models defined using TBModels\cite{Z2pack}.} 

\acknowledgements{}
Nordita is supported in part by NordForsk.

\bibliography{ref.bib}
\end{document}